\documentclass[aps,%
pra,%
twocolumn,%
superscriptaddress,%
longbibliography,%
reprint,%
]{revtex4-2}

\usepackage[english]{babel}
\usepackage[utf8]{inputenc}

\usepackage{graphicx}
\usepackage{amsmath, amssymb}
\usepackage{bm}
\usepackage{color}
\usepackage{xpatch}
\usepackage{natbib}
\usepackage{mathrsfs}
\usepackage{booktabs}
\usepackage{multirow}
\usepackage{array}

\makeatletter
\let\ORIbbl@fixname\bbl@fixname
\def\bbl@fixname#1{%
\@ifundefined{languagealias@\expandafter\string#1}
{\ORIbbl@fixname#1}
{\edef\languagename{\@nameuse{languagealias@#1}}}%
}
\newcommand{\definelanguagealias}[2]{%
\@namedef{languagealias@#1}{#2}%
}
\makeatother

\definelanguagealias{en}{english}
\definelanguagealias{en-US}{english}
\definelanguagealias{zh}{chinese}

\newcommand{\dd}{\,\mathrm{d}}
\newcommand{\mi}{\mathrm{i}}
\newcommand{\gagg}{g_{a\gamma\gamma}}

\begin{document}
\title{Coherently Enhanced Axion-Photon Conversion via Seeded Photons for Short-Pulse Axion Detection}

\author{Xiangyan An}
\affiliation{State Key Laboratory of Dark Matter Physics, Tsung-Dao Lee Institute, Shanghai Jiao Tong University, Shanghai 201210, China}
\affiliation{State Key Laboratory of Dark Matter Physics, Key Laboratory for Laser Plasmas (MoE), School of Physics and Astronomy, Shanghai Jiao Tong University, Shanghai 200240, China}
\affiliation{Collaborative Innovation Center of IFSA, Shanghai Jiao Tong University, Shanghai 200240, China}

\author{Min Chen}
\email{minchen@sjtu.edu.cn}
\affiliation{State Key Laboratory of Dark Matter Physics, Key Laboratory for Laser Plasmas (MoE), School of Physics and Astronomy, Shanghai Jiao Tong University, Shanghai 200240, China}
\affiliation{Collaborative Innovation Center of IFSA, Shanghai Jiao Tong University, Shanghai 200240, China}

\author{Jianglai Liu}
\email{jianglai.liu@sjtu.edu.cn}
\affiliation{State Key Laboratory of Dark Matter Physics, Tsung-Dao Lee Institute, Shanghai Jiao Tong University, Shanghai 201210, China}

\author{Yipeng Wu}
\affiliation{State Key Laboratory of Dark Matter Physics, Tsung-Dao Lee Institute, Shanghai Jiao Tong University, Shanghai 201210, China}

\author{Peng Yuan}
\affiliation{State Key Laboratory of Dark Matter Physics, Tsung-Dao Lee Institute, Shanghai Jiao Tong University, Shanghai 201210, China}

\author{Wenchao Yan}
\affiliation{State Key Laboratory of Dark Matter Physics, Key Laboratory for Laser Plasmas (MoE), School of Physics and Astronomy, Shanghai Jiao Tong University, Shanghai 200240, China}
\affiliation{Collaborative Innovation Center of IFSA, Shanghai Jiao Tong University, Shanghai 200240, China}

\author{Boyuan Li}
\affiliation{State Key Laboratory of Dark Matter Physics, Key Laboratory for Laser Plasmas (MoE), School of Physics and Astronomy, Shanghai Jiao Tong University, Shanghai 200240, China}
\affiliation{Collaborative Innovation Center of IFSA, Shanghai Jiao Tong University, Shanghai 200240, China}

\author{Feng Liu}
\affiliation{State Key Laboratory of Dark Matter Physics, Key Laboratory for Laser Plasmas (MoE), School of Physics and Astronomy, Shanghai Jiao Tong University, Shanghai 200240, China}
\affiliation{Collaborative Innovation Center of IFSA, Shanghai Jiao Tong University, Shanghai 200240, China}

\author{Zhengming Sheng}
\affiliation{State Key Laboratory of Dark Matter Physics, Tsung-Dao Lee Institute, Shanghai Jiao Tong University, Shanghai 201210, China}
\affiliation{State Key Laboratory of Dark Matter Physics, Key Laboratory for Laser Plasmas (MoE), School of Physics and Astronomy, Shanghai Jiao Tong University, Shanghai 200240, China}
\affiliation{Collaborative Innovation Center of IFSA, Shanghai Jiao Tong University, Shanghai 200240, China}

\author{Jie Zhang}
\affiliation{State Key Laboratory of Dark Matter Physics, Tsung-Dao Lee Institute, Shanghai Jiao Tong University, Shanghai 201210, China}
\affiliation{State Key Laboratory of Dark Matter Physics, Key Laboratory for Laser Plasmas (MoE), School of Physics and Astronomy, Shanghai Jiao Tong University, Shanghai 200240, China}
\affiliation{Collaborative Innovation Center of IFSA, Shanghai Jiao Tong University, Shanghai 200240, China}

\begin{abstract}
We propose a {seeded axion-photon conversion} scheme to enhance the sensitivity of light-shining-through-a-wall (LSW) experiments for axion detection, where the axions are generated from short pulse lasers and the usual resonant cavity is not applicable. By injecting a weak, coherent seed electromagnetic (EM) field into the axion-photon conversion region, the axion-induced EM field can constructively interfere with the seed field, amplifying the number of regenerated photons to a level exceeding that of the unseeded scenario. %
We evaluate the expected signal enhancement, statistical limits from Poisson counting with seed fluctuations and background, and the potential improvement in coupling sensitivity. Compared to a standard LSW setup, the seeded scheme can achieve orders-of-magnitude higher photon yield per axion, potentially surpassing resonance-enhanced experiments in certain parameter regimes. This approach presents a promising pathway to extend the reach of laboratory axion searches, particularly in scenarios where the resonant cavities are impractical.
\end{abstract}

\maketitle

\section{Introduction}

Axions and axion-like particles (ALPs) are well-motivated hypothetical bosons, originally introduced to solve the strong CP problem~\cite{peccei_cp_1977,peccei_constraints_1977,wilczek_problem_1978,weinberg_new_1978,kim_axions_2010} or extend the standard model~\cite{svrcek_axions_2006,Gelmini:1980re,Bellazzini:2017neg,ema_flaxion_2017,Calibbi:2016hwq,Irastorza:2018dyq}. 
Through their two-photon coupling $\gagg$, they can be probed in laboratory experiments using high-power lasers and strong magnetic fields. In a typical light-shining-through-a-wall (LSW) experiment~\cite{ehret_new_2010,bahre_any_2013,chou_search_2008,afanasev_experimental_2008,pugnat_results_2008,robilliard_no_2007}, a high-intensity laser propagates through a strong magnetic field, converting some photons into axions.  These axions pass through an opaque wall, then reconvert into photons in a second magnet on the far side of the wall. The regenerated photons would be detected as a signal of axions. Such experiments have been carried out by collaborations like ALPS~\cite{ehret_new_2010,Bahre:2013ywa}, GammeV~\cite{chou_search_2008}, LIPSS~\cite{afanasev_experimental_2008}, OSQAR~\cite{pugnat_results_2008}, and BMV~\cite{robilliard_no_2007}, and have placed the most stringent laboratory limits on axion-like particle couplings, excluding photon-axion coupling strengths $\gagg$ down to the $10^{-7}$-$10^{-8}$\,GeV$^{-1}$ level. However, the sensitivity of single-pass LSW experiments is extremely limited by the tiny photon-axion conversion probability, so more advanced techniques are required to probe weaker couplings.

One successful enhancement is the use of high-finesse Fabry-Perot cavities to resonantly boost the fields in the production and regeneration regions. The ALPS-II experiment, for example, employs a power-build-up cavity on the laser side (raising the effective laser power to $\sim$150\,kW) and a matched regeneration cavity on the detection side to amplify the reconverted photon number by nearly 4 orders of magnitude~\cite{bahre_any_2013,sikivie_resonantly_2007,mueller_detailed_2009}. Together with an interferometric heterodyne readout, ALPS-II is designed to improve sensitivity to $g_{a\gamma\gamma}$ down to about $2\times10^{-11}\,\mathrm{GeV}^{-1}$ for axions with mass smaller than 0.1\,meV, a dramatic leap over earlier LSW experiments~\cite{wei_towards_2024}.

The rapid development of high power laser technology and laser plasma studies provided a much stronger background magnetic field, from hundreds of Tesla to $10^6$\,Tesla~\cite{esarey_physics_2009,shi_efficient_2023,song_dense_2022,xue_generation_2023}. Such strong fields make it possible to generate axions with intense laser pulses, with typical durations ranging from nanoseconds to femtoseconds. However, the axion generated from these intense laser pulses would also be short pulses, making it challenging to enhance the axion regeneration process with the resonant cavities.

Specifically suited to the wakefield-based axion generation scheme proposed in~\cite{an_situ_2025}, here we focus on enhancing the photon regeneration process via seeding.
In this paper, we study a new approach to amplify the regenerated photon signal in short-pulse axion experiments by introducing a coherent seed electromagnetic field in the regeneration region. The idea is to inject a coherently controlled laser pulse (the seed) into the second magnet simultaneously with the arrival of the axion pulse, such that any axion-induced electromagnetic field will interfere with the seed EM field. If the seed is tuned to the same frequency and phase as the would-be regenerated photons, the interference will be constructive and effectively boost the detectable photon signal. This is analogous to %
optical parametric amplification in nonlinear optics, where a weak seed signal can stimulate and coherently enhance the conversion from the pump to the signal in a nonlinear crystal. In our case, the seed photons play a similar role to the injected signal light,  the axion field acts as the pump, and the background field acts as the crystal.

A related but distinct approach was proposed by Beyer \textit{et al.}~\cite{beyer_light-shining-through-wall_2022}, who use a laser to drive axion-photon scattering and mitigate phase mismatch at higher axion masses. Their enhancement arises from three-wave mixing rather than coherent superposition. In contrast, we work in the low-mass regime and amplify regeneration by letting the axion-induced field interfere coherently with a weak, phase-controlled seed behind the wall.

\begin{figure}[htbp]
    \centering
    \includegraphics[width=0.9\columnwidth]{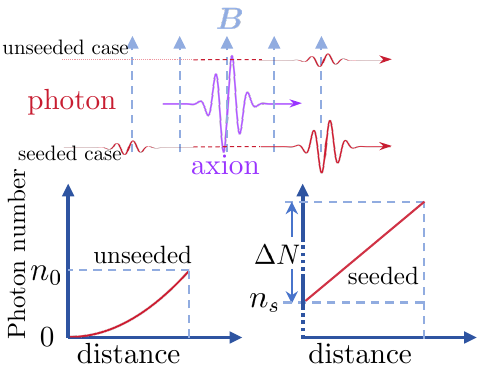}
    \caption{Conceptual scheme of coherently enhanced axion-photon conversion via seeded photons.}
    \label{fig:scheme}
\end{figure}

Figure~\ref{fig:scheme} shows the concept: Due to the smallness of $\gagg$, without a seed, the axion-to-photon reconversion yield is extremely small; when a weak, phase-controlled seed is injected, the axion-induced field interferes constructively with the seed, resulting in a larger regenerated photon number.

\section{Axion-Photon Conversion with a Seed Field}\label{sec:theory}
In the following, we quantitatively analyze the enhancement of the photon generation by a seed field. We first study the photon regeneration in the absence of a seed, then include a seed field in the process. Consider an axion field $\phi(t,\mathbf{x})$ coupled to the electromagnetic field. In the small mixing and weak-field regime, one can treat the axion-induced electromagnetic field as a perturbation $\mathbf{E}_1, \mathbf{B}_1$ on top of any applied background fields (the background being the static magnetic field $\mathbf{B}_0$ in the standard LSW scheme). The modified Maxwell equations are (in natural units $\hbar=c=\varepsilon_0=1$):
\begin{equation}
\nabla\times \mathbf{B}_1 - \partial_t \mathbf{E}_1 = g_{a\gamma\gamma}\,(\partial_t \phi)\,\mathbf{B}_0,
\end{equation}
along with $\nabla\cdot \mathbf{E}_1= -g_{a\gamma \gamma }\bm {B} \cdot \nabla \phi,$ $\nabla\cdot \mathbf{B}_1 = 0$ and $\nabla\times \mathbf{E}_1 + \partial_t \mathbf{B}_1 = 0$. From these, one obtains the wave equation of the vector potential $\mathbf{A}_1$ associated with the induced field:
\begin{equation}
\left(\partial_t^2 - \nabla^2\right)\mathbf{A}_1 = g_{a\gamma\gamma}\,(\partial_t \phi)\,\mathbf{B}_0\,,
\label{eq:wave}
\end{equation}
assuming the Coulomb gauge for $\mathbf{A}_1$. 

For a monochromatic axion source of angular frequency $\omega_0$ (with corresponding wave number $k_a = \sqrt{\omega_0^2 - m_a^2}$ for axion mass $m_a$), the axion field can be written as a plane wave $\phi(t,x) = \phi_0\,e^{\mi(k_a x - \omega_0 t)}$ propagating in the $+x$ direction (we take the wall to be at $x=0$ and the regeneration magnet spanning $0 < x < l_2$). We look for a solution of $\mathbf{A}_1$ with the form $\mathbf{A}_1(t,x) = \hat{\mathbf{e}}\,\mathcal{A}(x)\,e^{\mi(k_1 x - \omega_0 t)}$, where $k_1 = \omega_0$ is the wave number of a photon with the same frequency and $\hat{\mathbf{e}}$ is the polarization unit vector. Here $\mathcal{A}(x)$ is a slowly varying complex amplitude envelope of the wave in the regeneration region. Substituting it into Eq.~(\ref{eq:wave}) and assuming $\partial_x \mathcal{A} \ll k_1 \mathcal{A}$ (i.e. the field amplitude $\mathcal{A}$ varies little over a wavelength), we obtain a first-order differential equation:
\begin{equation}
2\mi k_1\,\frac{\dd\mathcal{A}(x)}{\dd x} = \mi g_{a\gamma\gamma}\,\omega_0\,\phi_0\,B_0\,e^{\mi(k_1 - k_a)x}\,. \label{eq:diff}
\end{equation}
The difference $k_1 - k_a = \kappa$ represents the momentum mismatch between the axion and photon waves. For $m_a \ll \omega_0$ (nearly massless axion), $\kappa \approx \frac{m_a^2}{2\omega_0}$; if $m_a = 0$, then $\kappa = 0$ meaning a perfect phase matching. Integrating Eq.~(\ref{eq:diff}) from $0$ to $x$ yields the solution of the photon field amplitude at some position $x$ in the regeneration region:
\begin{equation}
\mathcal{A}(x) = \mathcal{A}(0) + \frac{g_{a\gamma\gamma}\, \omega_0\, \phi_0\, B_0}{2\mi k_1}\left[\frac{e^{\mi \kappa\,x}-1}{\kappa}\right]\!. \label{eq:A_solution}
\end{equation}
The first term is the homogeneous solution, representing the evolution of any initial field $\mathcal{A}(0)$ present at the entrance of the region ($x=0^+$, immediately behind the wall). The second term is the particular solution sourced by the axion. In a standard LSW experiment with no seed photons behind the wall, one has the boundary condition $\mathcal{A}(0)=0$. In that case, Eq.~(\ref{eq:A_solution}) reduces to the usual result: the regenerated photon field amplitude $\mathcal{A}_\text{axion}(l_2)$ is proportional to $g_{a\gamma\gamma} B_0 l_2$ (for small $\kappa\,l_2$), and the probability for an axion to convert into a photon is 
\begin{equation}
P_{a\to\gamma} = \frac{|\mathcal{A}_{\text{axion}}(l_2)|^2}{|\phi_0|^2} \approx \frac{1}{4}g_{a\gamma\gamma}^2 B_0^2 l_2^2\,\mathcal{S}(m_a)^2\,,
\label{eq:P_std}
\end{equation}
where $\mathcal{S}(m_a) = \frac{\sin(\kappa\,l_2/2)}{\kappa\,l_2/2}$ is a suppressing factor for nonzero $m_a$ arising from phase matching (for $m_a=0$, $\kappa=0$, $\mathcal{S}=1$). If $N_a$ axions enter the regeneration region, the expected number of photons produced is simply $N_\gamma = N_a\,P_{a\to\gamma}$.

Now consider the scenario with a seed field. We prepare an initial electromagnetic field $\mathcal{A}(0)\neq 0$ behind the wall, consisting of $n_s$ photons at frequency $\omega_0$ (we assume the seed field co-propagates collinearly with the axion wave and has a polarization parallel to $\mathbf{B}_0$ to maximize axion-photon conversion). This seed field could, for instance, be derived from a split-off portion of the same laser used to generate the axions with an appropriate time delay, ensuring it enters the regeneration magnet coincidentally with the axion pulse. The presence of $\mathcal{A}(0)$ modifies the solution for $\mathcal{A}(x)$ to Eq.~(\ref{eq:A_solution}).

The total intensity of the electromagnetic wave at $x=l_2$ is proportional to $|\mathcal{A}(l_2)|^2$. To obtain the number of photons, we integrate the intensity over the interaction volume. The result can be written as 
\begin{equation}
N'_\gamma = \omega_0 \int_{V}  |\mathbf{A}_1|^2 \dd V \propto |\mathcal{A}(l_2)|^2\,,
\end{equation}
up to normalization constants that cancel out when comparing with the unseeded case. Using Eq.~(\ref{eq:A_solution}) and expanding the magnitude squared, we find
\begin{equation}
N'_\gamma = |\mathcal{A}(0)|^2 + 2\,\mathrm{Re}\!\big[\mathcal{A}(0)^* \mathcal{A}_{\text{axion}}(l_2)\big] + |\mathcal{A}_{\text{axion}}(l_2)|^2\,,
\end{equation}
where $\mathcal{A}_{\text{axion}}(l_2)$ denotes the particular solution (the second term in Eq.~(\ref{eq:A_solution})) evaluated at $x=l_2$. Identifying $n_s = |\mathcal{A}(0)|^2$ as the number of seed photons and $n_0 = |\mathcal{A}_{\text{axion}}(l_2)|^2$ as the number of photons that would be regenerated in the absence of a seed (i.e. $n_0 = N_a\,P_{a\to\gamma}$ from Eq.~(\ref{eq:P_std})), the above expression becomes 
\begin{equation}
N'_\gamma = n_s + n_0 + 2\sqrt{n_s\,n_0}\cos\delta\,,
\label{eq:N_with_seed}
\end{equation}
where $\delta$ is the relative phase between the seed field and the axion-induced field at $x=l_2$. In the optimal case, the experiment is configured such that the axion-induced photons emerge in phase with the seed ($\delta=0$, i.e. constructive interference). In that case, the interference term is maximized and we have 
\begin{equation}
N'_\gamma = n_s + n_0 + 2\sqrt{n_s\,n_0}\,.
\label{eq:Nprime}
\end{equation}
This result can be intuitively understood: the axion converts into a photon that is coherently added to an existing electromagnetic field. The $2\sqrt{n_s n_0}$ term represents the cross-term where one photon arises from the axion converted field and one from the seed field. Equation~(\ref{eq:Nprime}) serves as the core formula governing the enhancement of seeded regeneration.

We are often interested in the net generated photons attributable to axions. Since $n_s$ seed photons were injected, one may subtract the expected contribution from the seed field and define the increase in the number of signal photons as 
\begin{equation}
\Delta N \equiv N'_\gamma - n_s = n_0 + 2\sqrt{n_s\,n_0}\,.
\label{eq:DeltaN}
\end{equation}
In the absence of any seed ($n_s=0$), $\Delta N$ correctly reduces to $n_0$. But for $n_s \gg n_0$, $\Delta N \approx 2\sqrt{n_s\,n_0} \gg n_0$. The seed field thus amplifies the detectable photon number increase by a factor 
\begin{equation}
\mathcal{E} \equiv \frac{\Delta N}{n_0} = 1 + 2\sqrt{\frac{n_s}{\,n_0\,}}\,.
\label{eq:enhancement_factor}
\end{equation}
For $n_s \gg n_0$, this simplifies to $\mathcal{E} \approx 2\sqrt{n_s/n_0}$. In other words, the fractional increase in photon signal scales as the square root of the seed photon number (assuming the seed is strong compared to the original signal). %
Nonetheless, since $n_0$ is extremely small in practice, even a modest $n_s\sim 1$ can yield a very large $\sqrt{n_s/n_0}$ factor.

Equation~(\ref{eq:enhancement_factor}) indicates that the benefit of seeding is greatest when $n_s \gg n_0$. %
For instance, consider a $100$\,J, $800$\,nm pulse (about $4\times10^{20}$ photons) traversing a total generation field integral $B_1 l_1=3000\,\text{T}\cdot\text{m}$. For $g_{a\gamma\gamma}=10^{-9}\,\mathrm{GeV}^{-1}$ this gives a total number of produced axions of $N_a \approx 8.8 \times10^{8}$. With a $B_2 l_2 = 500\,\text{T}\cdot\text{m}$ regeneration region and $m_a\approx0$, Eq.~(\ref{eq:P_std}) yields $P_{a\to\gamma}\approx 6.1 \times 10^{-14}$, so the unseeded expectation per shot is $n_0=N_a P_{a\to\gamma}\approx 5.4 \times 10^{-5} \ll 1$. Even a single seed photon ($n_s = 1$) would produce an interference term $\Delta N\approx 2\sqrt{n_0n_s}\approx 1.4\times 10^{-2}$, corresponding to an enhancement by about three orders of magnitude over the unseeded case.

\section{Statistical Case Study}\label{sec:enhancement}

For a realistic seed source, the number of seed photons may fluctuate from shot to shot, so an additional measurement is required to determine the seed photon number $n_s$.
To illustrate how such fluctuations influence the attainable sensitivity, we consider three
representative cases summarized in Table~\ref{tab:cases}. In all cases, $N_{\text{shot}}$ denotes the total number of laser shots 
and $n_b$ denotes the mean photon number originating from background.

\begin{table}[htbp]
  \centering
  \renewcommand{\arraystretch}{1.2}
  \setlength{\tabcolsep}{6pt}
  \newcolumntype{P}{>{\centering\arraybackslash}m{12mm}}
  \begin{tabular}{cccc}
    \toprule
    &  & \textbf{Signal} & \textbf{Background} \\
    \midrule
  Case~1 & & $N_{\text{shot}}\cdot n_{0}$ & $N_{\text{shot}}\cdot n_{b}$ \\[2pt]
    \midrule
  \multirow{2}{*}{Case~2}
      & ON  & $\frac{1}{2}N_{\text{shot}}\!\left(n_{s}+n_{0}+2\sqrt{n_{0}n_{s}}\right)$ & $\frac{1}{2}N_{\text{shot}}\cdot n_{b}$ \\[2pt]
      & OFF & $\frac{1}{2}N_{\text{shot}}\cdot n_{s}$ & $\frac{1}{2}N_{\text{shot}}\cdot n_{b}$ \\[2pt]
    \midrule
  Case~3 &  & $N_{\text{shot}}\cdot (n_0+2\sqrt{n_{0}n_{s}})$ & $N_{\text{shot}}\cdot n_{b}$ \\
    \bottomrule
  \end{tabular}
  \caption{Summary of the statistical configurations used to evaluate
the median sensitivity on the axion-induced photon number $n_0$.
 \label{tab:cases}}
\end{table}

{Case~1:} This case represents the baseline situation without any seeded photons. 
The measured counts follow a Poisson distribution with mean $N_{\text{shot}} (n_b + n_0)$. 
The median sensitivity at 90\% confidence level (CL)~\footnote{The median sensitivity is obtained by simulating large number of background-only experiments and taking the mean of the upper limits of the signal at 90\% from these simulations.} on $n_0$ is evaluated using the Feldman-Cousins unified approach~\cite{feldman_unified_1998}, which constructs a confidence interval based on likelihood-ratio ordering for Poisson counting with a known background. 

{Case~2:} Seed photons are introduced but their number fluctuates from shot to shot with a Poisson distribution. We have no knowledge of $n_s$ at all, therefore have to determine it statistically.
To achieve this, the measurement is divided into an ``ON'' run (axion conversion enabled) and an ``OFF'' run (conversion disabled), both using $N_{\text{shot}}/2$ laser shots. 
The OFF run measures only the seed and background photons, giving an expectation $\tfrac{1}{2}N_{\text{shot}}(n_s+n_b)$, 
while the ON run contains the same background plus the axion-induced term, $\tfrac{1}{2}N_{\text{shot}}\!\left(n_s+n_b+n_0+2\sqrt{n_0n_s}\right)$. 
The difference between these two runs isolates the coherent axion contribution $n_0+2\sqrt{n_0n_s}$, while naturally accounting for the statistical uncertainty in $n_s$. 
Because both ON/OFF counts fluctuate independently according to Poisson statistics, 
the median sensitivity on $n_0$ is obtained using the likelihood-ratio method of Rolke, Lopez, and Conrad~\cite{rolke_limits_2005}, 
which extends the Feldman--Cousins formalism to include an auxiliary background measurement with finite statistics. 

{Case~3:} Finally, in an idealized situation where the seed photon number is fully known and reproducible.
The measurement reduces to a single Poisson process with an expected signal of $N_{\text{shot}}(n_0+2\sqrt{n_0n_s})$ and the same background $N_{\text{shot}}n_b$.
The median sensitivity on $n_0$ is again evaluated following the
Feldman-Cousins prescription. 

\begin{figure}[tb]
    \centering
  \includegraphics[width=0.99\columnwidth]{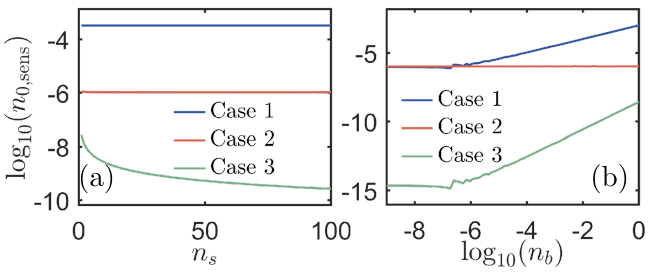}
  \caption{(a) Median sensitivity (sens) on $n_0$ at 90\% CL versus $n_s$ at fixed $N_{\text{shot}}=30\,\text{days}\times 1\,\text{Hz}$ and $n_b=0.1$. (b) Same sensitivities versus $n_b$ at fixed $n_s=100$ and $N_{\text{shot}}=30\,\text{days}\times 1\,\text{Hz}$. All sensitivities use the statistical configurations in Table~\ref{tab:cases}. \label{fig:enhancement_n0ul}}
\end{figure}

The corresponding results are shown in Fig.~\ref{fig:enhancement_n0ul}. 
To provide analytical insight, we consider the large-count limit for each configuration.
For case~2, under the approximation $n_s \gg n_b,\,N_\text{shot}n_s \gg 1$, 
when both runs yield the same mean background level $\tfrac{1}{2}N_{\text{shot}}n_s$, 
the Rolke method yields an upper limit on the signal of approximately $1.64\sqrt{N_{\text{shot}}n_s}$, where $\sqrt{N_{\text{shot}}n_s}$ reflects the Poisson fluctuation of the seed counts and the factor $1.64$ corresponds to the 90\% confidence level. In this regime, the ON/OFF difference targets the effective signal, $\tfrac{1}{2}N_{\text{shot}}(n_0+2\sqrt{n_0n_s}) \approx N_{\text{shot}}\sqrt{n_0n_s}$, implying an upper limit (UL) on $n_0$ of $n_{0,\text{UL}} \approx 1.64^2/N_{\text{shot}}$ that is essentially independent of $n_s$. 
For case~1, when the background is significant ($N_{\text{shot}}n_b \gg 1$), the upper limit on the signal is about $1.64\sqrt{N_{\text{shot}}n_b}$ while the target signal is $N_{\text{shot}}n_0$, hence $n_{0,\text{UL}} \approx 1.64 \sqrt{n_b/N_{\text{shot}}}$. 
In this situation, although case~2 requires ON/OFF measurements, it can still outperform case~1 to some extent. 
However, when the background is very low ($N_{\text{shot}}n_b \ll 1$), the Feldman-Cousins method yields an upper limit of about 2.4 events on the signal ($N_{\text{shot}}n_0$) at 90\% CL. Thus case~1 gives $n_{0,\text{UL}} \approx 2.4/N_{\text{shot}}$. 
In this regime, the limits from case~1 and case~2 become nearly identical. These analytical upper-limit scalings describe the asymptotic behavior of
the expected median sensitivity shown in Fig.~\ref{fig:enhancement_n0ul}.
In all scenarios, when the seed photon number is perfectly known (case~3), the coherent enhancement term $2\sqrt{n_0n_s}$ yields the lowest attainable upper limit on $n_0$.
These comparisons show that the precision in characterizing the seed photon number directly determines how effectively the coherent enhancement can be utilized in setting the axion-photon coupling limit.

It is also important to note that strict shot-to-shot reproducibility of the seed is not essential: what matters is that we know the photon number in each shot with sufficient certainty. Given that the initial photon number of the seed field can be measured with high precision, then the variation between different shots can be corrected for in the data analysis. This may be done by using correlated photon sources, such as pair production processes (e.g. laser induced double photon emission process), where one photon of the pair can be used to monitor the seed photon count, and the other one is injected as the actual seed. In such a scheme, the seed photon fluctuations can be strongly reduced, because the per-shot photon number is tagged by the reference measurement.

Another important consideration is the phase alignment between the seed and axion-induced field. In writing Eq.~(\ref{eq:Nprime}) we assumed perfect constructive interference ($\cos\delta = 1$). If the phase $\delta$ were instead $\pi$ (180$^\circ$ out of phase), the interference term would be negative, leading to {destructive} interference where $N'_\gamma = n_s + n_0 - 2\sqrt{n_s n_0}$. In that case the presence of axions would {deplete} the seed beam slightly, reducing the photon count at the detector. This scenario is not actually detrimental to detection---one could still measure a difference $\Delta N$ (which would now be negative) by comparing, as long as the magnitude $|\Delta N| = 2\sqrt{n_s n_0}$ is large enough. Nevertheless, to maximize signal, the experimenters would ideally synchronize the seed phase such that $\delta \approx 0$. In practice, the axion field phase is set by the phase of the initial laser in the production region and any phase shift accumulated in transit. If the seed is split from the same laser source, one can maintain a fixed phase relationship via interferometric stability over the beam paths. For a pulsed system, locking the relative phase may be challenging, but since the effect depends on $\cos\delta$, even a slight phase error only reduces the interference term by a cosine factor. Incoherent averaging over random $\delta$ would in the worst case eliminate the $2\sqrt{n_s n_0}$ term on average (since positive and negative contributions would cancel), so phase control is highly preferable to realize the seeded enhancement.

\section{Implications for Sensitivity}\label{sec:discussion}
The ultimate goal of an LSW experiment is to detect or constrain the axion-photon coupling $g_{a\gamma\gamma}$.
We therefore translate the photon-counting observables into the median sensitivity on $g_{a\gamma\gamma}$, explicitly incorporating the per-shot background $n_b$ in the sensitivity estimate.

An important advantage of using short-pulse, ultra-intense laser drivers for axion generation and detection 
is that the extremely short temporal window of the interaction effectively suppresses the number of 
background photons per laser shot. For a typical dark-count rate of about $10~\mathrm{Hz}$ in a
single-photon detector, axion-induced photons are expected to appear only within the laser pulse duration 
of tens of femtoseconds. Consequently, the background photon number per shot can be suppressed down 
to the level of $n_b\sim 10^{-15}$. Even when taking into account the detector's timing gate (typically of 
nanosecond duration), the effective background photon number remains as low as $n_b\sim 10^{-8}$ per shot. Therefore, the background $N_\text{shot}n_b$ is negligible even with $N_\text{shot}=30\,\text{days}\times1\,\text{Hz}=2.6\times 10^{7}$. Under such low-background conditions, 
cases 1 and 2 yield nearly identical median sensitivities on $n_0$ (and thus on $g_{a\gamma\gamma}$). 
In the following discussion, we mainly compare cases 1 and 3. 
For a null-result experiment, the Feldman-Cousins method gives a $90\%$ CL upper limit of about 
2.4 signal photons, which we use to estimate the median sensitivity on $g_{a\gamma\gamma}$.

For a standard unseeded LSW experiment, the median sensitivity on $g_{a\gamma\gamma}$ is given by $N_\text{shot}n_0 =2.4$.
Writing $N_{\text{shot}}=R\,T_{\text{run}}$ (with $R$ the repetition rate of the pulses and $T_{\text{run}}$ the total running time), we have $g_{a\gamma\gamma}$ sensitivity (in case of no detection) by:
\begin{equation}
\gagg^\text{sens} \approx 2.5\left({R T_\text{run}n_L}\right)^{-1/4}\left(B_1l_1B_2l_2\,\mathcal{S}_1\mathcal{S}_2\right)^{-1/2},
\label{eq:g_no_seed}
\end{equation}
where $n_L$ is the number of photons per laser shot (related to laser pulse energy $P_L \tau_L / (\hbar \omega_0)$), $B_1, l_1$ are the generation magnet field and length, $B_2, l_2$ for regeneration, and $\mathcal{S}_{1,2}$ are phase mismatch suppression factors for each stage. 
Equation\,(\ref{eq:g_no_seed}) essentially comes from $n_0 = \frac{1}{4} g_{a\gamma\gamma}^4 B_1^2 l_1^2 B_2^2 l_2^2 n_L$ and the requirement $R T_{\text{run}}\,n_0 =2.4$. 
Plugging in some representative numbers, where axions are produced in the generation stage via a laser-wakefield interaction~\cite{an_situ_2025}
and detected through an LSW setup (without resonance in the regeneration cavity): $R=1\,\text{Hz}$, $T_{\text{run}}=30$\,days, $n_L=4\times10^{20}$ (100\,J at 800\,nm), $B_1 l_1 = 3000\,\text{T}\cdot\text{m}$, $B_2 l_2 = 500\,\text{T}\cdot\text{m}$, and assuming near-optimal $\mathcal{S}_1 \approx \mathcal{S}_2 \approx 1$, we find 
\begin{equation}
\gagg^\text{sens} \approx 4.3\times 10^{-10}\,\mathrm{GeV}^{-1}\,,
\end{equation}
which is indeed weaker sensitivity than ALPS-II (by almost an order of magnitude). The lack of resonant buildup on both the generation and regeneration cavities contributes to this relatively high floor on $g_{a\gamma\gamma}$.

\begin{figure}
\centering
\includegraphics[width=0.7\columnwidth]{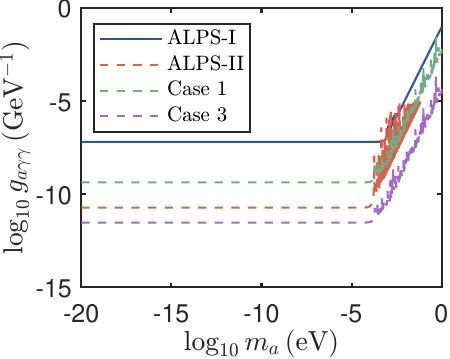}
\caption{Projected sensitivity to $g_{a\gamma\gamma}$ for a pulsed LSW experiment with and without seeding, compared to ALPS experiment~\cite{bahre_any_2013,AxionLimits}. \label{fig:constraint}}
\end{figure}

In the seeded experiment, the criterion for detection changes: we require $R T_{\text{run}}\,\Delta N = 2.4$ (same upper limit at 90\% CL from Feldman-Cousins method). Using $\Delta N \approx 2\sqrt{n_0 n_s}$ (valid if $n_s \gg n_0$) and noting $n_0 = g_{a\gamma\gamma}^4\,C$ where $C = \frac{1}{16}n_L B_1^2 l_1^2 B_2^2 l_2^2$ collects all the experimental constants, we get:
\begin{equation}
  \gagg^\text{sens} \approx 2.2\left({R^2 T_\text{run}^2n_Ln_s}\right)^{-1/4}\left(B_1l_1B_2l_2\,\mathcal{S}_1\mathcal{S}_2\right)^{-1/2}.
\label{eq:g_with_seed}
\end{equation}
Comparing with Eq.~(\ref{eq:g_no_seed}), we see that effectively the factor $R\,T_{\text{run}}$ is under a square root instead of a fourth root, and $n_s$ also appears under a fourth root in the denominator. This implies that long integration times and high repetition rates are {more powerful} in improving sensitivity for the seeded case than in the unseeded case. 
Inserting the same numerical values as above and taking $n_s=100$ photons, Eq.~(\ref{eq:g_with_seed}) gives 
\begin{equation}
g_{a\gamma\gamma}^\text{sens} \approx 3.0\times 10^{-12}\,\mathrm{GeV}^{-1}\,,
\end{equation}
for the seeded experiment.
Figure~\ref{fig:constraint} illustrates this comparison: in the unseeded, single-pass configuration our projected limit on $g_{a\gamma\gamma}$ remains weaker than the ALPS-II sensitivity; by contrast, introducing a modest seed of $n_s=100$ photons per shot coherently enhances the regeneration and pushes the reach below the ALPS-II constraint.
Of course, this dramatic gain assumes ideal interference and negligible additional noise beyond shot noise. It shows, nonetheless, that seeded regeneration can partially compensate for the lack of resonant enhancement in pulsed experiments.

\section{Conclusion}\label{sec:conclusion}
We have proposed a seeded photon regeneration scheme for axion searches with high-power laser pulses. We show that by injecting a coherent seed electromagnetic field into the regeneration region, the coherent interaction between the axion field and the photons can enhance the axion-induced photon signal by orders of magnitude, thereby improve the experimental sensitivity to the axion-photon coupling. The method circumvents the incompatibility of resonant cavities with femtosecond pulses, operates effectively in the time domain, and remains advantageous even under realistic noise constraints, provided that seed intensity fluctuations are monitored via independent means. This approach motivates further studies and has the potential to significantly extend the reach for axions through the pulsed light-shining-through-a-wall.

\section*{Acknowledgments}
\label{sec-acknowledgement}

This work was supported by the National Natural Science Foundation of China (Grants No. 12225505, 12090060, 12090061), Office of Science and Technology and Shanghai
 Municipal Government (Grant No. 23JC1410200), Shanghai Jiao Tong University 2030 Initiative, and partially supported by State Key Laboratory of Dark Matter Physics. We thank for the sponsorship from the Hongwen Foundation in Hong Kong and New Cornerstone Science Foundation in China. The computations in this paper were run on the $\pi$ 2.0 cluster supported by the Center for High Performance Computing at Shanghai Jiao Tong University.

\bibliography{seeded_axion_regeneration}

\end{document}